\documentclass[11pt]{article}

\usepackage{fullpage}
\usepackage{setspace}
\usepackage{parskip}
\usepackage{titlesec}
	\titlespacing{\section}{0pt}{*3}{*1}
	\titleformat{\section}
		{\sffamily\bfseries\large\uppercase}  	% FORMAT
		{\thesection.}{.5em}{}[]							% LABEL	% SEP %  BEFORE SECTION NAME % AFTER													
	\titlespacing{\subsection}{0pt}{*2}{*1}
	\titleformat{\subsection}
		{\sffamily\bfseries}  	% FORMAT
		{\thesubsection.}{.5em}{}[]							% LABEL	% SEP %  BEFORE SECTION NAME % AFTER		
	\titlespacing{\subsubsection}{0pt}{*1.5}{*0.5}
	\titleformat{\subsubsection}
		{\sffamily\itshape}  	% FORMAT
		{\thesubsubsection.}{.5em}{}[]							% LABEL	% SEP %  BEFORE SECTION NAME % AFTER	
\usepackage[section]{placeins}
\usepackage{xcolor}
\usepackage{breakcites}
\usepackage{lineno}
\usepackage{hyphenat}
\usepackage{times}

\renewenvironment{abstract}
  {{\bfseries\noindent{\abstractname}\par\nobreak}\normalsize}
  {\bigskip}

\titlespacing{\section}{0pt}{*3}{*1.5}
\titlespacing{\subsection}{0pt}{*2}{*1}
\titlespacing{\subsubsection}{0pt}{*1.5}{*0.5}

\usepackage{authblk}
\usepackage{graphicx}
\usepackage[outdir=./]{epstopdf}
\usepackage[space]{grffile}
\usepackage{latexsym}
\usepackage{textcomp}
\usepackage{longtable}
\usepackage{tabulary}
\usepackage{booktabs,array,multirow}
\usepackage{amsfonts,amsmath,amssymb}
\newif\iflatexml\latexmlfalse

\AtBeginDocument{
	\DeclareGraphicsExtensions{.pdf,.PDF,.eps,.EPS,.png,.PNG,.tif,.TIF,.jpg,.JPG,.jpeg,.JPEG}}

\usepackage[english]{babel}
\usepackage{float}
\usepackage[margin=1in]{geometry}
\usepackage[style=base]{caption}
\newtheorem{theorem}{Theorem}%
\graphicspath{{figures/}}
\DeclareMathAlphabet      {\ibd}{OML}{cmm}{b}{it}
\newcommand{\df}	{\mathrm{d}f}
\newcommand{\dg}	{\mathrm{d}g}
\newcommand{\dr}	{\mathrm{d}r}
\newcommand{\dmu}{\mathrm{d}\mu}
\newcommand{\eref}[1]{(\ref{#1})}
\newcommand{\fref}[1]{Fig.~\ref{#1}}
\makeatletter
\def\munderbar#1{\underline{\sbox\tw@{$#1$}\dp\tw@\z@\box\tw@}}
\makeatother

\begin{document}

\title{Balancing at the edge of excitability: Implications for cell movement}
\author[1,2]{Debojyoti Biswas} %
\author[3]{Parijat Banerjee}
\author[2,*]{Pablo A. Iglesias}
\affil[1]{Department of Mechanical Engineering, The Johns Hopkins
University, 3400 N. Charles St., Baltimore, MD 21218 United
States.}
\affil[2]{Department of Electrical \& Computer Engineering, The Johns
Hopkins University, 3400 N. Charles St., Baltimore, MD
21218 United States.}
\affil[3]{Department of Physics \& Astronomy, The Johns Hopkins
University, 3400 N. Charles St., Baltimore, MD 21218 United
States.}
\date{}                                           % Activate to display a given date or no date

\begingroup
\let\center\flushleft
\let\endcenter\endflushleft
\maketitle
\endgroup

\abstract{Cells rely on the ability to sense and respond to small spatial differences in chemoattractant concentrations for survival.   There is growing evidence that this is accomplished by setting the signaling system near the threshold for activation in an excitable system and using the spatial heterogeneities to alter the threshold thereby biasing cell activity in the direction of the gradient.  Here we consider a scheme by which the set point is adaptively set near the bifurcation point, but without explicit knowledge of this point. Through simulation, we show that the method would improve chemotactic efficiency of cells.  The results of this paper are based on pioneering work by Eduardo Sontag and  coworkers, to whom this paper is dedicated in honor of his 70th birthday.     }


\maketitle\newpage
%-------------------------------------------------------------------------------------------------
\section{Introduction}\label{sec:intro}
%-------------------------------------------------------------------------------------------------
Many biological systems operate near the edge of a threshold.  In some cases, these are used as a means of filtering noise. An example is the activation of T-cells, white blood cells that form an integral part of the adaptive immune system~\cite{Guram:2019aa}.  T-cell activation is triggered through the binding of the T-cell receptor by specific antigens from pathogens.  This can eventually lead to a cytotoxic response, by which ``killer T-cells'' dispose of virus-infected cells.  A key component of this pathogen recognition is the ability to adjust the threshold of activation through feedback loops that rearrange the receptors spatially~\cite{Bene:2020aa}. A similar effect is present in natural killer cells~\cite{Narni-Mancinelli:2013aa}.  

Adaptive thresholds play an important in hearing.  The exquisitely sensitivity by which humans can detect sound comes, in part, from amplification of the received signals~\cite{Hudspeth:2014aa}.  This amplification also relies on an adaptation mechanism.  Hair cell bundles that sense physical sounds and trigger the adaptive response can oscillate spontaneously as well as in response to these external stimuli.  The oscillatory nature of the hair bundle behavior has been proposed to originate in the presence of a Hopf bifurcation~\cite{Choe:1998un,Camalet:2000uy, Eguiluz:2000ui}.  The dynamical equilibrium is assumed to lie on the stable side but some distance away from of a Hopf bifurcation.  This distance acts as threshold --- subthreshold stimuli elicit no response.  On the other hand, as the threshold diminishes, hearing becomes progressively more sensitive. Of course, once the threshold is crossed, the loss of stability leads to autonomous oscillations, which are undesirable and, in hearing,  may represent tinnitus~\cite{Jackson:2019tj}.  

In our work, we study \emph{chemotaxis} --- the motion of cells based on external gradients of chemoattractants --- of large eukaryotic cells.  Unlike fast moving bacteria that interpret spatial gradients by moving quickly in a medium and differentiating the receptor occupancy signal over time, eukaryotic cells are large and slow and use a spatial mechanism to sense the gradient~\cite{Parent:1999aa}.  Cells of the  model organism \emph{Dictyostelium discoideum}  display remarkable sensitivity to these gradients: a difference in receptor occupancy of only five receptors between front and back can guide their motion~\cite{Haastert:2007aa}.  In \emph{Dictyostelium}, the amplification of this small spatial asymmetry is achieved through an excitable network~\cite{Vicker:2002ab,Xiong:2010aa}.
  
Excitable networks, first used by Hodgkin and Huxley to describe the signaling of neurons, represent an important class of dynamical systems in biology \cite{Hodgkin:1948aa}.  Typically, the system operates at a stable equilibrium.  Small external stimuli or internal stochastic fluctuations are largely filtered out. However, for a sufficiently large stimulus, a stereotypical large-scale excursion in phase-space occurs before the system returns to its steady state.  This threshold-like behavior allows the cell to filter our noise.   In practice, the size of the threshold determines the likelihood of a response \cite{Bhattacharya:2018aa}.  There is evidence that cells modulate the threshold of the signaling system in response to the external chemoattractant gradient, lowering it at the front and raising it at the rear so as to bias cell movement towards the gradient.  We recently showed experimentally that altering this threshold leads to highly oscillatory cells  in both \emph{Dictyostelium} cells and various mammalian cell lines \cite{Miao:2017aa}. Moreover, this altered threshold appears to be a hallmark of metastatic progression in epithelial cancer models \cite{Zhan:2020aa}.  Interestingly, the resultant signal from the excitable network is then fed to a second excitable network that triggers the actual motion of the cell. This cytoskeletal subsystem shows characteristics of an excitable system that is poised near a Hopf bifurcation \cite{Westendorf:2013aa}. Thus, chemotaxis involves a coupled system of excitable networks \cite{Huang:2013aa,Miao:2019aa}.  

These examples illustrate the importance of setting the operating point and corresponding threshold.  However, in the uncertain environment of cell physiology, this requires highly adaptable control systems.   In a series of articles, Sontag and co-workers presented a set of rules for ensuring that a dynamical system can operate in the vicinity of a bifurcation point~\cite{Moreau:2003aa,Moreau:2003ab}.  Since first coming across these papers, we have been intrigued by the possibility of using the techniques as a means of improving chemotactic efficiency.  Here, we revisit these results and, in particular, use them to show how they could be used to enhance the chemotactic behavior of cells that rely on an excitable system to move.
%-------------------------------------------------------------------------------------------------
\section{Results}\label{sec2}
%-------------------------------------------------------------------------------------------------

%-------------------------------------------------------------------------------------------------
\subsection{Preliminaries}
\label{sec:prelim}
%-------------------------------------------------------------------------------------------------
To set the stage for the results that follow, we use the following system, originally considered by Moreau et al. \cite{Moreau:2003ab}.
\begin{subequations}
\label{eq:Sontag}
\begin{align}
	\dot{x}&=y\\
	\dot{y}&=-(\mu_0-\mu)y-\lambda{y}^3-\omega^2{x}.
\end{align}
\end{subequations}
When $\mu<\mu_0$, the origin represents a stable equilibrium, with the system undergoing a Hopf bifurcation at $\mu=\mu_0$;  oscillations exist when $\mu>\mu_0$. Henceforth we seek a means of driving $\mu$ to its bifurcation point $\mu_0$ with the proviso that $\mu_0$ is unknown.  In particular, we seek an adaptation law  of the form
\begin{equation}
	\label{eq:Adapt1}
	\dot{\mu} = f(x,y)-g(\mu)
\end{equation}
In the case where $\lambda=0$, Moreau et al. \cite{Moreau:2003ab} showed that, if
\begin{equation}
	\label{eq:SontagA} 
	f(x,y)=-\tfrac{1}{2}a\ln(x^2+(y/\omega)^2),\qquad
	g(\mu)=b\mu
\end{equation}
then the adaptation~\eref{eq:Adapt1} drives the system to a unique periodic orbit which is globally stable and that $\mu=\mu_0$. The result in which we are most interested, however, is the following.  Here, let $r=\sqrt{x^2+(y/\omega)^2}$.
\begin{theorem}[\cite{Moreau:2003ab}]
Let $\mu_0\in\mathbf{R}$ and consider continuously differentiable functions $f:\mathbf{R}_{>0}\rightarrow\mathbf{R}$ and $g:\mathbf{R}\rightarrow\mathbf{R}$. Assume that $g(\mu_0)$ is in the image of $f$. Consider $r^\star\in f^{-1}(g(\mu_0))$ and assume that $0<-(\df/\dr)(r^\star)r^\star\le((\dg/\dmu)(\mu_0))^2$ and
$(\dg/\dmu)(\mu_0)>0$. Then the system of equations \eref{eq:Sontag} with $\lambda=0$ and \eref{eq:SontagA} with $(x,y)\in\mathbf{R}^2\backslash\{(0,0)\}$ and $\mu\in\mathbf{R}$ has a periodic orbit that is locally exponentially stable and where $\mu=\mu_0$.
\end{theorem}

%-------------------------------------------------------------------------------------------------
\begin{figure}[tp!]
\centering
\includegraphics[width=1\textwidth]{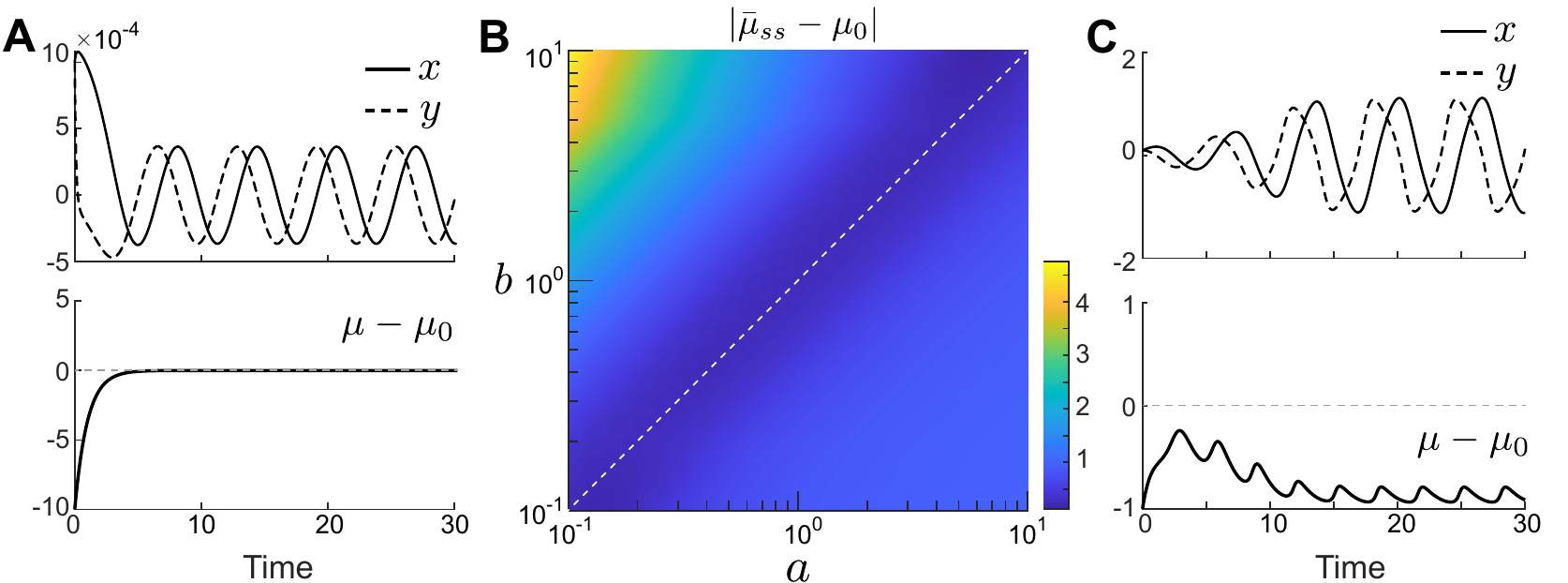}
\caption{\label{fig1}Bifurcation control of nonlinear oscillator. %
(A) Simulation of \eref{eq:Sontag} with update law: $\dot{\mu} = a/(1+x^2+y^2/\omega^2)-b\mu$ \cite{Moreau:2003aa}. Parameter values used are: $a = b =\mu_0=\lambda = \omega = 1$. %
(B) Effect of variation of $a,b$ on the steady state error in $\mu$ ($\lvert \bar{\mu}_{ss}-\mu_0\lvert$). The white dashed line in B indicates zero error. The parameter space corresponds to the right side of the dashed line ($a\geq b$) results in the oscillatory response. %
(C) Simulation of \eref{eq:Sontag} with a different update law: $\dot{\mu} = a/(1+25x^2+y^2/\omega^2)-b\mu$
with same parameter values used in panel A, drives the system to a different value than $\mu_0$.}
\end{figure}
%-------------------------------------------------------------------------------------------------
An important aspect of this result is the notion that if the system is going to approach an equilibrium  after the implementation of the adaptation law with $\mu=\mu_0$, then there must be a value $r^\star$ such that
\begin{equation}
\label{match_param}
	f(x^\star,y^\star)=g(\mu_0),
\end{equation}
where $x^\star$ and $y^\star$ are the points such that $r^\star=\sqrt{(x^\star)^2+(y^\star/\omega)^2}$. Clearly, if $x^\star$ and $y^\star$ are the equilibrium values, then $f(x^\star,y^\star)$ is constant.  However, for $f(x^\star,y^\star)$ to be constant,  $x^\star$ and $y^\star$ need not be.  This is clearly possible in the case of \eref{eq:Sontag} with $\lambda=0$  where the energy of the corresponding harmonic oscillator
\[
	E=\tfrac{1}{2}\left(x^2+(y/\omega)^2\right),
\]
is constant during the oscillation.  This, however, will not be true in general,  as when $\lambda\ne0$.  In the latter case, however, it can be shown that  for sufficiently small perturbations the system will be driven to a point whereby $\mu$ approaches $\mu_0$ \cite{Moreau:2003ab}.  

The form of the adaptation law~\eref{eq:SontagA} is not unique~\cite{Moreau:2003aa}. An alternate choice  is
\[
	f(x,y) = \frac{a}{1+x^2+y^2}
\]
  also works in this context, as illustrated in \fref{fig1}A.

We note that the requirement of \eref{match_param} is a crucial one as it determines how close $\mu$ can approach $\mu_0$ and this, in turn, depends on  accurate choices of parameters in the update law ($a,b$) as well as a constant value function $f(x^\star,y^\star)$ at the Hopf bifurcation. As shown in \fref{fig1}B, choosing values of $a$ and $b$ for which \eref{match_param}  does not hold means that $\vert\mu-\mu_0\vert\not\rightarrow0$. Similarly, using an incorrect $f$ can lead to significant loss of performance (\fref{fig1}C).

In this present study we are interested in designing similar update laws for excitable systems and analyzing the effects of  perturbations particularly as they apply to bifurcations. 
%-------------------------------------------------------------------------------------------------
\subsection{An excitable system}
\label{sec:FHN}
%-------------------------------------------------------------------------------------------------
In 1961, Richard FitzHugh suggested a mathematically tractable simplification of the original Hodgkin-Huxley equations~\cite{FitzHugh:1961aa}.  Independently, Nagumo et al. designed and built an electrical circuit that recreated these dynamics.  The  FitzHugh-Nagumo (FHN) is a special class of two variable systems with a fast and a slow state variables.  In describing excitable behavior, we assume that the system is operating at a stable equilibrium that lies close to a Hopf bifurcation. When this point is crossed, the system exhibits relaxation oscillations.

\begin{figure}[tp!]
\centering
\includegraphics[width=.95\textwidth]{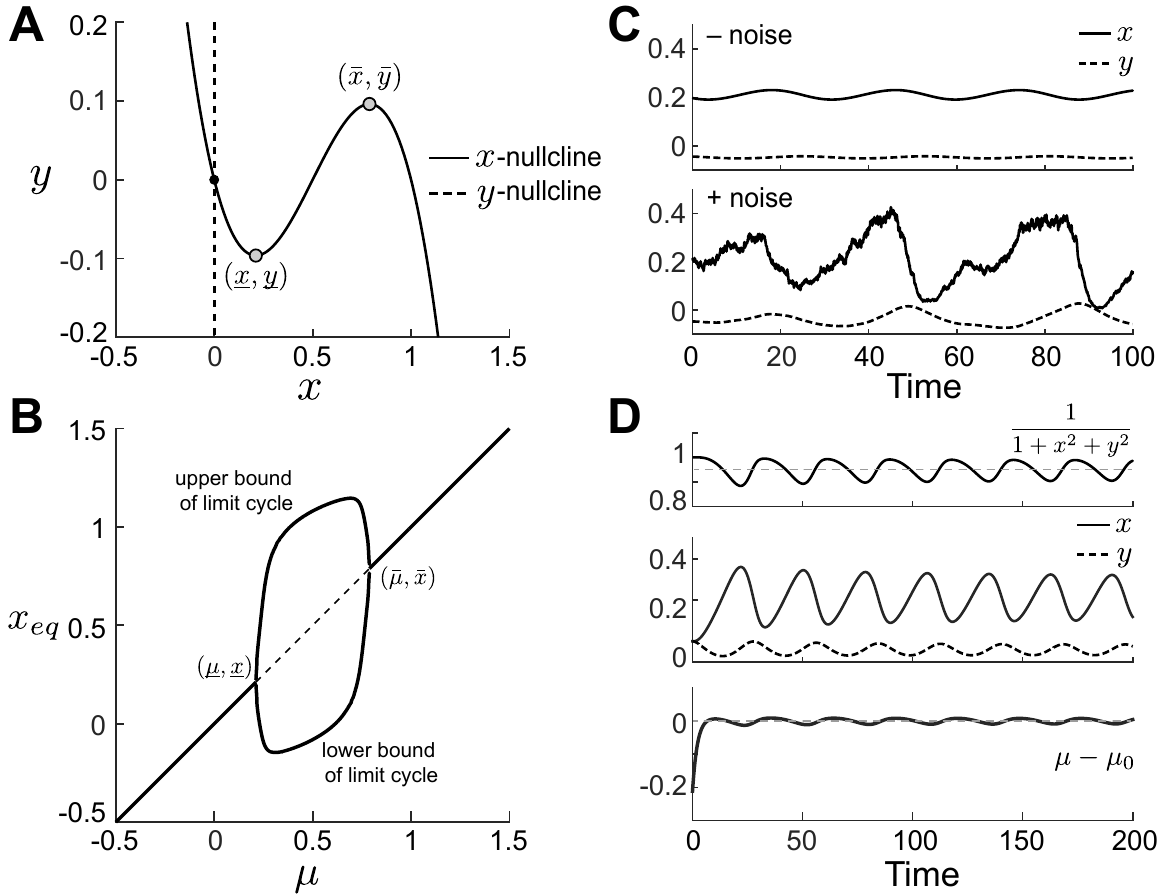}
\caption{\label{fig2}  FHN model. 
(A) $x$ and $y$-nullclines with Hopf bifurcation points shown as the grey circles. %
(B) Bifurcation diagram showing $x_{eq}$ as the function of parameter $\mu$. The dashed section corresponds to the oscillatory response. %
(C) Temporal responses of $x$ and $y$ in absence (top) and presence (bottom) of noise. The parameter values used are: $\lambda = 0.5$, $\varepsilon = 0.05$ and $w = \mathcal{N}(0,0.5)$. %
(D) Temporal responses of $f(x,y) = 1/(1+x^2+y^2),x,y$ and the error ($\mu-\mu_0$) for the FHN system coupled with the $\mu$-update law: $\dot{\mu} = \varepsilon_{\mu}(af(x,y)-b\mu)$, in absence of any noise. The parameter values used for the update law: $a=1,b = 0.95/\underline{\mu},\varepsilon_{\mu} = 0.1$.
}
\end{figure}

We consider the following set of equations:
\begin{subequations}
\begin{align}
	\dot{x}&=-x(x-\lambda)(x-1)-y+w,\qquad 0<\lambda<1\\
	\dot{y}&=\varepsilon(x-\mu)
\end{align}
\end{subequations}
where $x$ and $y$ are  fast and slow variables, respectively, and $w$ is an external Gaussian noise input with mean zero and variance $\sigma^2$.  In the absence of noise, the fast variable has a cubic nullcline (\fref{fig2}A):
\[
	y=h(x)=-x(x-\lambda)(x-1).
\]
The nullcline for the slow variable is a vertical line; together, they lead to a unique equilibrium $(x,y)_{eq}=(\mu,h(\mu))$ that is  asymptotically stable if either $\mu<\munderbar{\mu}$ or $\mu>\bar{\mu}$, where 
$\munderbar{\mu}$ and $\bar{\mu}$ are the local minimum and maximum of the function $h(x)$.  When $h^\prime(\mu)=0$, the system undergoes Hopf bifurcations (\fref{fig2}B).  If $\lambda=\frac{1}{2}$, these occur at $\munderbar{\mu}\approx 0.211$ and $\bar{\mu}\approx0.789$, and the corresponding state variables are
\[
	(\munderbar{x},\munderbar{y})\approx(0.211,-0.048),\quad
	(\bar{x},\bar{y})\approx(0.789,0.048).
\]
When operating at one of the Hopf bifurcation points the system exhibits sustained small amplitude oscillations.  In the presence of noise, these oscillations are replaced by occasional large amplitude responses typical of  excitable dynamics (\fref{fig2}C).

%-------------------------------------------------------------------------------------------------
%-------------------------------------------------------------------------------------------------

We now use the procedure described in Section~\ref{sec:prelim} to create an adaptation law that can drive the system to one of the Hopf bifurcation points.  We focus on the one at the lower end of the $x$ variable (\munderbar{x}) corresponding to $\munderbar{\mu}$.  Define the adaptation law according to \eref{eq:Adapt1} with
\begin{equation}
\label{naive_adapt}
   f(x,y)=\frac{1}{1+x^2+y^2}=\frac{1}{1+r^2},\quad g(\mu) = b\mu.
\end{equation}
We see that unlike the system \eref{eq:Sontag}, $f(x,y)$ is not constant, but oscillates around $\bar{f} \approx 0.95$ (\fref{fig2}D).  To satisfy \eref{match_param}, we use this mean level of $f$: $\bar{f}$ and this leads to the choice for $b=\bar{f}/\underline{\mu}\approx 4.5$ in \eref{naive_adapt}. Owing to the oscillations in $f(x,y)$, we also observed oscillations in the error $\mu-\mu_0$. 

Eliminating the error in $\mu$, requires designing a function $f$ that is constant at values during the oscillation. Thus, $f(r^\star)$ is a conserved quantity in the system (such as the energy in the harmonic oscillator).  In a nonconservative system,  such a quantity does not exist. However, when a system changes its stability through Hopf bifurcation, the resulting oscillation will trace a closed curve in the phase plane. Suppose that we  approximate the resulting limit cycle by an ellipse using, for example, the Levenberg-Marquardt (LM) method~\cite{Gill:1978aa}
\begin{equation}
	\label{eq:ellipse}
	\mathcal{C^\star} 
	= \mathcal{C}_A(x^\star)^2
	+\mathcal{C}_Bx^\star y^\star 
	+\mathcal{C}_C(y^\star)^2
	+\mathcal{C}_Dx^\star 
	+\mathcal{C}_Ey^\star
	+\mathcal{C}_F
\end{equation}
and that we implement an update law given by:
\begin{equation}
\label{conic_update}
\dot{\mu} = \varepsilon_{\mu}\bigg(\dfrac{1}{1+\mathcal{C}^\star}-b\mu\bigg)
\end{equation}

%-------------------------------------------------------------------------------------------------
\begin{figure}[tp!]
\centering
\includegraphics[width=.95\textwidth]{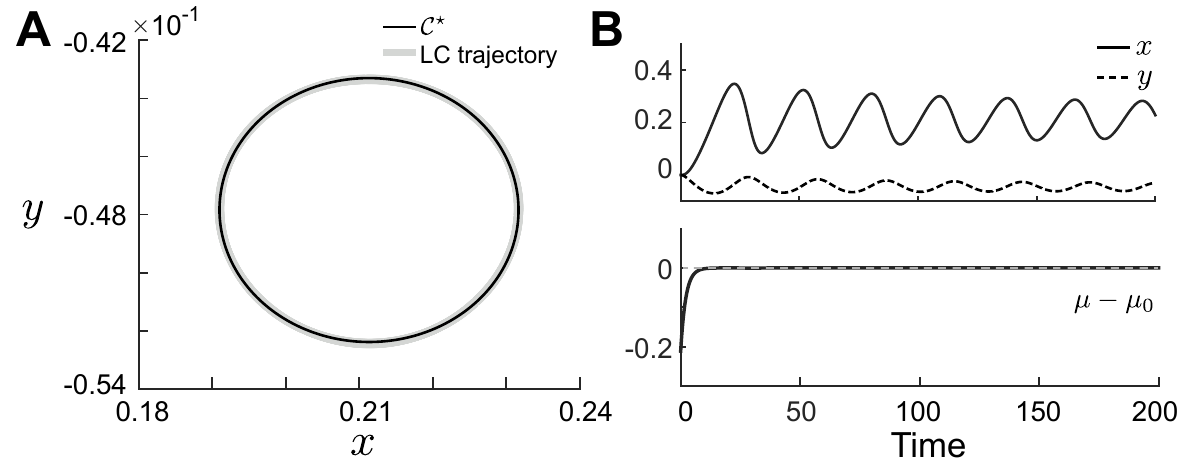}
\caption{\label{fig3}  Precise bifurcation control of FHN model. 
(A) Limit cycle trajectory (grey) and the best fitting ellipse \eref{eq:ellipse}, 
%$\mathcal{C}^\star=\mathcal{C}_Ax^2+\mathcal{C}_Bxy+\mathcal{C}_Cy^2+\mathcal{C}_Dx+\mathcal{C}_Ey+\mathcal{C}_F$ (black) 
correspond to $\underline{\mu}\approx 0.211$. %
(B) Temporal responses of $x$, $y$ and the error ($\mu-\mu_0$) corresponds to the $\mu$-update law with $f(x,y)=1/(1+\mathcal{C}^\star)$. The parameter values used for the update law: 
$a=1$, 
$b = 1/\underline{\mu}$,
$\varepsilon_{\mu} = 0.1$,
$\mathcal{C}_A \approx 4.5\times10^{-2}$, 
$\mathcal{C}_B \approx 4.1\times 10^{-4}$, 
$\mathcal{C}_C \approx 0.99$, 
$\mathcal{C}_D \approx -2.1\times 10^{-2}$, 
$\mathcal{C}_E \approx 9.5\times 10^{-2}$ and 
$\mathcal{C}_F \approx 4.5\times 10^{-3}$.}
\end{figure}
%-------------------------------------------------------------------------------------------------

Though the system states are time varying quantities during the oscillation,  depending on the goodness of its fit,  $\mathcal{C}^*$ will be close to being constant and hence so will $f(x,y) = 1/(1+\mathcal{C}^*)$. In \fref{fig3}, we illustrate the effectiveness of such a function on the excitable FHN system. As shown in \fref{fig3}A, the approximation is quite good when the system is operating close to the bifurcation.  The resulting system is oscillatory and $\mu$ approaches $\mu_0$ (\fref{fig3}B).

\begin{figure}[htp!]
\centering
\includegraphics[width=.95\textwidth]{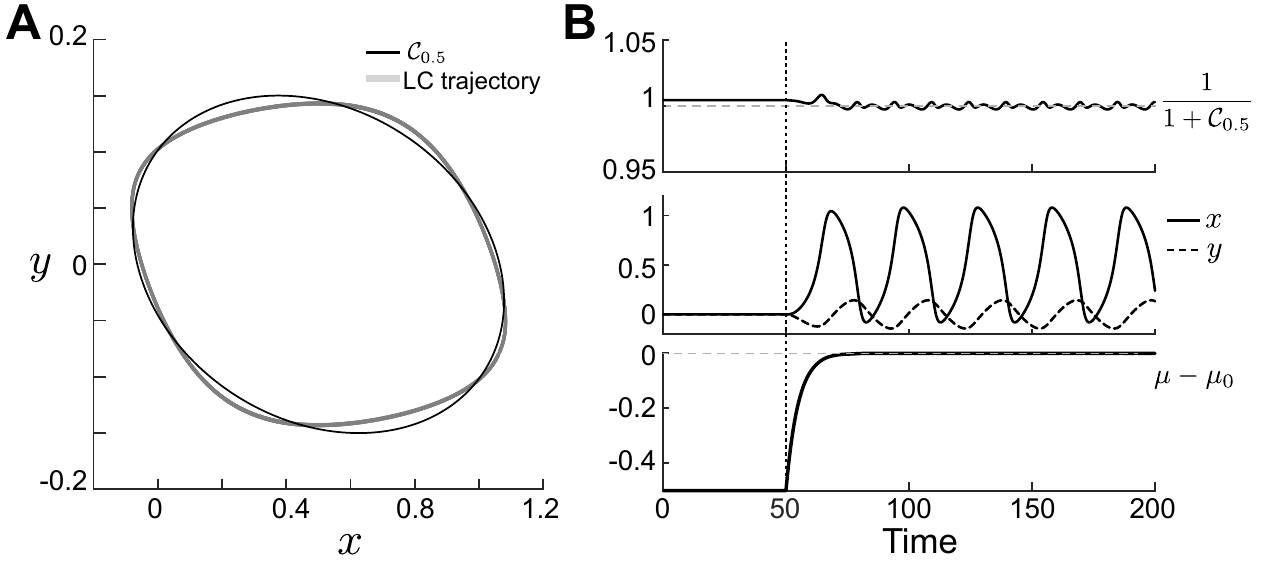}
\caption{\label{fig4}  Control of FHN model in the oscillatory regime. 
(A) Limit cycle trajectory (grey) and the best fitting ellipse, $\mathcal{C}_{0.5}$ 
% =Ax^2+Bxy+Cy^2+Dx+Ey+F$ 
(black) correspond to $\mu_0= 0.5$. 
(B) Temporal responses of $f(x,y)=1/(1+\mathcal{C}_{0.5}),x,y$ and the error ($\mu-\mu_0$). The parameter values used for the update law: 
$a=1$, 
$b = 2$, 
$\varepsilon_{\mu} = 0.1$, 
$\mathcal{C}_A \approx 6.7\times10^{-2}$, 
$\mathcal{C}_B \approx 0.11\times 10^{-4}$, 
$\mathcal{C}_C \approx 0.99$, 
$\mathcal{C}_D \approx -6.7\times10^{-2}$, 
$\mathcal{C}_E = -5.6\times10^{-2}$ and
$\mathcal{C}_F \approx 4.5\times 10^{-3}$. The update law was switched on at $t = 50$, denoted by the black dotted line. 
}
\end{figure}

The application of the such update law is not restricted  to the control near the bifurcation point, but can also be applied to drive the system to any desired setting. The design task involves finding appropriate $a$, $b$ and $f(x,y)$. For a fixed equilibrium point it is trivial. For operating in the oscillatory regime, we could follow similar step from \eref{conic_update} to construct an ellipse, $C_\mu$ enclosing the trajectories on the phase space. This is illustrated in \fref{fig4}, where the system is oscillating at $\mu = 0.5$  far from either bifurcation point.  Note that, in this case, the best fitting ellipse obtained by the LM method is distinguishable from the actual limit cycle (\fref{fig4}A) and hence the system does not settle to a constant $f(x^\star,y^\star)$ (\fref{fig4}B); nevertheless, the bifurcation parameter $\mu$ does approach $\mu_0$.
%-------------------------------------------------------------------------------------------------
%-------------------------------------------------------------------------------------------------

%-------------------------------------------------------------------------------------------------
\subsection{The signaling excitable system regulating chemotaxis}
\begin{figure}[tp!]
\centering
\includegraphics[width=.95\textwidth]{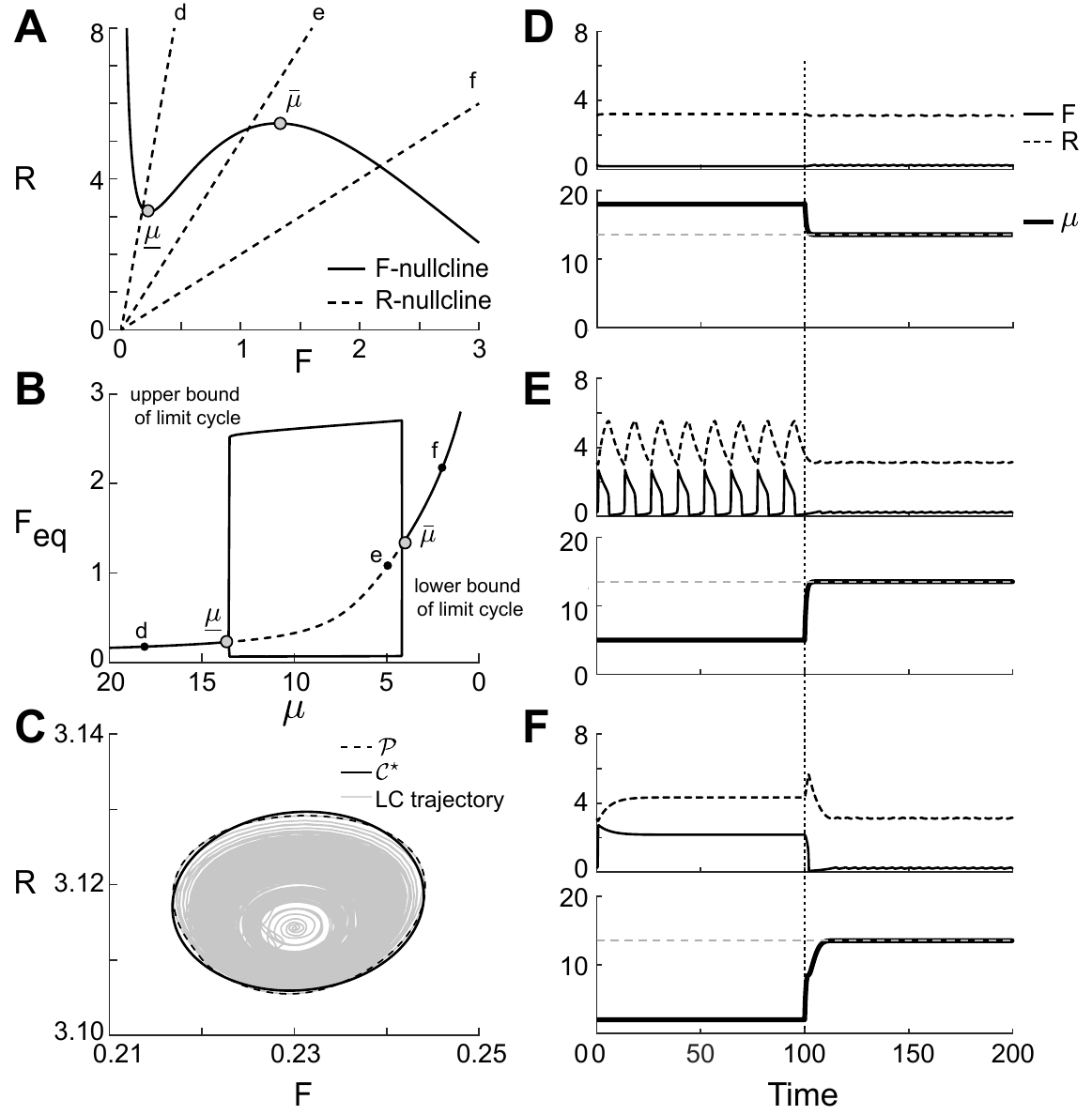}
\caption{\label{fig5}  Bifurcation control of F-R model. 
(A) $F$- and $R$-nullclines with Hopf bifurcation points ($\munderbar{\mu},\bar{\mu}$) shown as the grey circles. Three $R$-nullclines denoted as d, e and f correspond to the parameters $\mu_d>\mu_e>\mu_f$.
(B) Bifurcation diagram showing $F_{eq}$ as the function of parameter $\mu$. The dashed section corresponds to the oscillatory response.
(C) Limit cycle trajectory (grey), convex hull $\mathcal{P}$ (black dashed) and the best fitting ellipse, $\mathcal{C}^\star$ (black solid) correspond to $\munderbar{\mu}$. 
(D-F) Temporal responses of $F$, $R$ and $\mu$ for different initial conditions of $\mu$: $\mu_d=18$ (D), $\mu_e = 5$, (E), $\mu_f = 2$ (F). The parameter values used in the simulation are: 
$a_1=0.083$, 
$a_2 = 8.33$, 
$a_3 = 93.7$, 
$a_4 = 2.4$, 
$a_5 = 0.735$, 
$a_6 = 4$, 
$\varepsilon = 0.1$, 
$ \varepsilon_{\mu} = 20$, 
$\mu_0 = \munderbar{\mu} \approx 13.535$, 
$k = 1/\mu_0$,  
$\mathcal{C}_A \approx 6.6\times10^{-2}$, 
$\mathcal{C}_B \approx -1\times 10^{-2}$, 
$\mathcal{C}_c \approx 8.6, \mathcal{C}_D \approx 1.3\times10^{-3}, \mathcal{C}_E = -0.54, \mathcal{C}_F= 0.84$. The update law was switched on at $t = 100$ denoted by the black dotted line. }
\end{figure}

%-------------------------------------------------------------------------------------------------
In the context of biochemical signaling in which states represent concentration of various species, the FHN system is not realizable, as it allows the state variables to be negative.  Through a series of papers \cite{Xiong:2010aa, Shi:2013aa, Biswas:2021ab}, we proposed a biochemically plausible model of the excitable system regulating cell motility and validated it experimentally~\cite{Miao:2019aa}. Consider the following set of equations.  
\begin{subequations}
\label{eq:EN}
\begin{align}
	\dot{F}&=-(a_1+a_2R)F+\bigg(\frac{a_3F^2}{a_4^2+F^2}+a_5\bigg)(a_6-F)+w\\
	\dot{R}&=\varepsilon(\mu F-R)
\end{align}
\end{subequations}
The variables $F$ and $R$ refer to fast and refractory states. The terms involving $a_1$ and $a_2$ represent degradation of the fast variable, with the latter being a part of a negative feedback loop. The terms involving $a_3$ and $a_4$ are part of a positive feedback loop on $F$ that saturates and has cooperativity with Hill coefficient of two. The equation for $R$ shows linear activation of the refractory state, initiating the negative feedback loop, and a constant degradation rate. 
 For the time being we will ignore the noise, $w$ in the system. The nullcline for the fast variable retains the ``inverted N" shape of the FHN system, and that of the slow variable is  linear with slope $\mu$ (\fref{fig5}A). Note that the slope of this line acts as the bifurcation variable (\fref{fig5}B). Whereas steep slopes lead to a unique stable steady state with low values for the two states,  shallow slopes lead to a permanently high stable equilibrium. Between these two extremes, oscillatory behavior is possible.  As above, the two bifurcation points are denoted by $\underline{\mu}$ and $\overline{\mu}$, respectively. Note, however, that unlike the FHN model, it is the steeper of the two slopes ($\underline{\mu}$) that results in the lower level of activity and so $\underline{\mu}>\overline{\mu}$ (\fref{fig5}A).

%-------------------------------------------------------------------------------------------------
%-------------------------------------------------------------------------------------------------

Following the form of Section~\ref{sec:FHN}, we designed a control law to drive the system \eref{eq:EN} to the bifurcation point corresponding to the lower $F$ concentration:
\begin{equation}
\label{eq:EN_adapt}
	\dot{\mu} =\varepsilon_{\mu}\bigg(\dfrac{1}{1+\mathcal{C}^\star}-k\mu\bigg) 
\end{equation}
%where $\mathcal{C}^\star$ is the best fitted ellipse enclosing the limit cycle trajectories at Hopf bifurcation point. 
As there is no closed form solution of  \eref{eq:EN} to determine the limit cycle trajectory, we relied on  numerical simulations. Because  the simulations near the Hopf bifurcation point are quite sensitive to  numerical error, resulting in oscillations of varying amplitude and frequency, we constructed a convex polygon, $\mathcal{P}$, containing all the trajectories over a sufficiently long time (1000 time units) and then computed the ellipse of best fit,  $\mathcal{C}^\star$, to $\mathcal{P}$ (\fref{fig5}C). We then tested whether the update law \eref{eq:EN_adapt} was capable of driving the system to the intended bifurcation point 
from three characteristically different operating regimes: %
(1) low $F$ and $R$ states correspond to $\mu>\munderbar{\mu}$; %
(2) oscillating $F$ and $R$ states correspond to $\munderbar{\mu}>\mu>\bar{\mu}$; and %
(3) high $F$ and $R$ states correspond to $\bar{\mu}>\mu$.
In all three simulations, $\mu$ approached the bifurcation point.  When approaching it from below (\fref{fig5}D), the change in state was minimal as the system settled to its new operating point, which is close to the initial state.  In the case where the initial condition was in the oscillatory regime (\fref{fig5}E), these oscillations disappeared quickly.  Finally, when approaching the new set point from above  (\fref{fig5}F), the system underwent one ``firing'' before settling to the desired equilibrium point.  

%-------------------------------------------------------------------------------------------------
\begin{figure}[tp!] %[tp!]
\centering
\includegraphics[width=.95\textwidth]{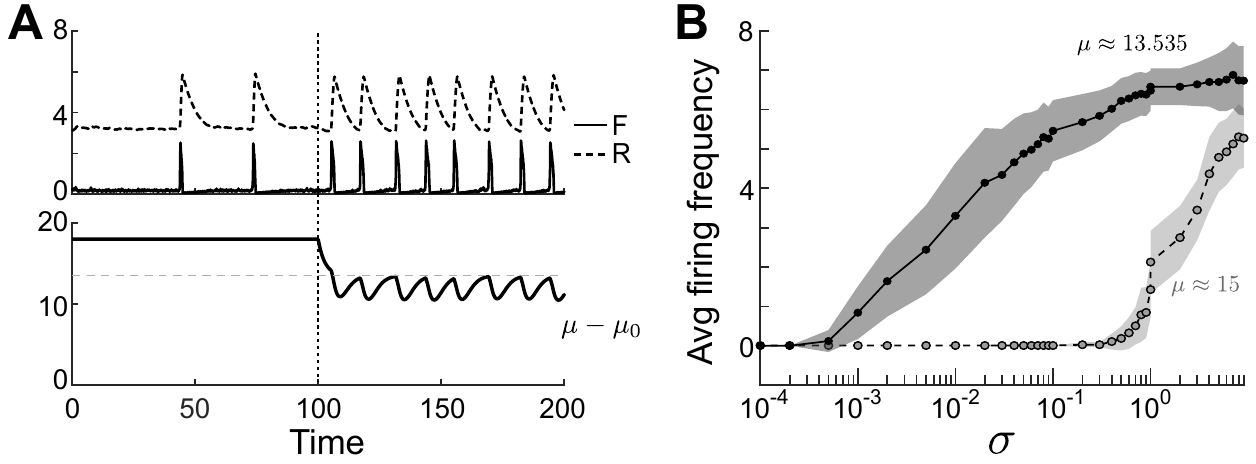}
\caption{\label{fig6}  Bifurcation control of F-R model in presence of  noise. 
(A) Temporal responses of $F$ and $R$ and the error ($\mu-\mu_0$). The update law was switched on at $t = 100$ denoted by the black dotted line. The noise parameter used is $\sigma = 10$.
(B) Average firing frequency in  a time window of 100 A.U. for $\mu \approx 13.535$ (black markers, Hopf bifurcation value) and  $\mu = 15$ (grey markers). The respective shaded regions denote the standard deviations  for 10 independent simulations.}
\end{figure}
%-------------------------------------------------------------------------------------------------

So far, we have ignored the role of noise. However, in the excitable system paradigm of of cell movement, noise plays an important role as it is stochastic perturbations that actually lead to firings and subsequent cellular protrusions~\cite{Xiong:2010aa}. Thus, one might expect that operating at or close to the bifurcation point might actually lead to undesirable behavior as noise would continuously trigger excitable waves.  To examine this, we simulated the system in the presence of noise with or without the adaptation law (\fref{fig6}A).  As can be seen in the simulation shown in \fref{fig6}A, noise leads to occasional firings. However, after the adaptation law is turned on and the system settles close to the bifurcation point, the barrier for having noise trigger the system is essentially eliminated. This results in continuous firings that appear to be periodic. It should be emphasized, however, that these are not the small scale oscillations that would be predicted by the Hopf bifurcation theory, but are rather large scale excursions that are characteristic of excitable systems. Note, however, that the time between them is essentially eliminated, but that the period is determined by the deterministic properties of the system. This \emph{autonomous stochastic resonance} phenomenon has been studied for a number of excitable systems~\cite{Lindner:2004aa}.  We note that, as shown in \fref{fig6}B, operating some distance away from the bifurcation point leads to more threshold-like stochastic triggering of activity.  When the operating point was set to $\mu=15$, firings were only triggered when the size of the noise variance approached $\sigma=1$.  In contrast, when setting  $\mu=\munderbar{\mu}$,  firings appears almost immediately, and their number increased almost linearly with the noise variance.

%-------------------------------------------------------------------------------------------------
\subsection{Effect on chemotaxis}
\label{sec:chemotax}
%-------------------------------------------------------------------------------------------------
The simulations so far show that bringing the system towards its bifurcation point can lead to an increase in the number of firings and hence a more active cell.  If this happens in a spatially heterogeneous manner, however, the increased activity is likely to impair chemotaxis. Experimentally, this has been demonstrated~\cite{Miao:2017aa}. Global recruitment of an enzyme to the cell membrane led to a lowering of the threshold to the point where the entire cell started oscillating synchronously. This increase in activity, however, resulted in no net movement. To increase \emph{directed} cell motility, the effect of lowering the threshold has to be spatially confined to the regions of high chemoattractant.  

We simulated the effect of applying the update law to a cell exposed to a spatial gradient in the two dimensional reaction-diffusion F-R model.  The equations are as follows:
\begin{subequations}
\label{eq:FRdiff}
\begin{align}
	\frac{\partial F}{\partial t}&=-(a_1+a_2R)F+\bigg(\frac{a_3F^2}{a_4^2+F^2}+a_5\bigg)(a_6-F)+w+D_F\frac{\partial^2 F}{\partial \theta^2}\\
	\frac{\partial R}{\partial t}&=\varepsilon(\mu F-R)+D_R \frac{\partial^2 R}{\partial \theta^2} \label{eq:EN_diffusion}
\end{align}
\end{subequations}
where we have included diffusion (the last terms of each equation) of the respective species. The spatial domain, parameterized by $\theta$ is assumed to  be one-dimensional  with periodic boundary conditions (suggestive of a two-dimensional cell membrane). 

%-------------------------------------------------------------------------------------------------
\begin{figure}[tp!]
\centering
\includegraphics[width=1\textwidth]{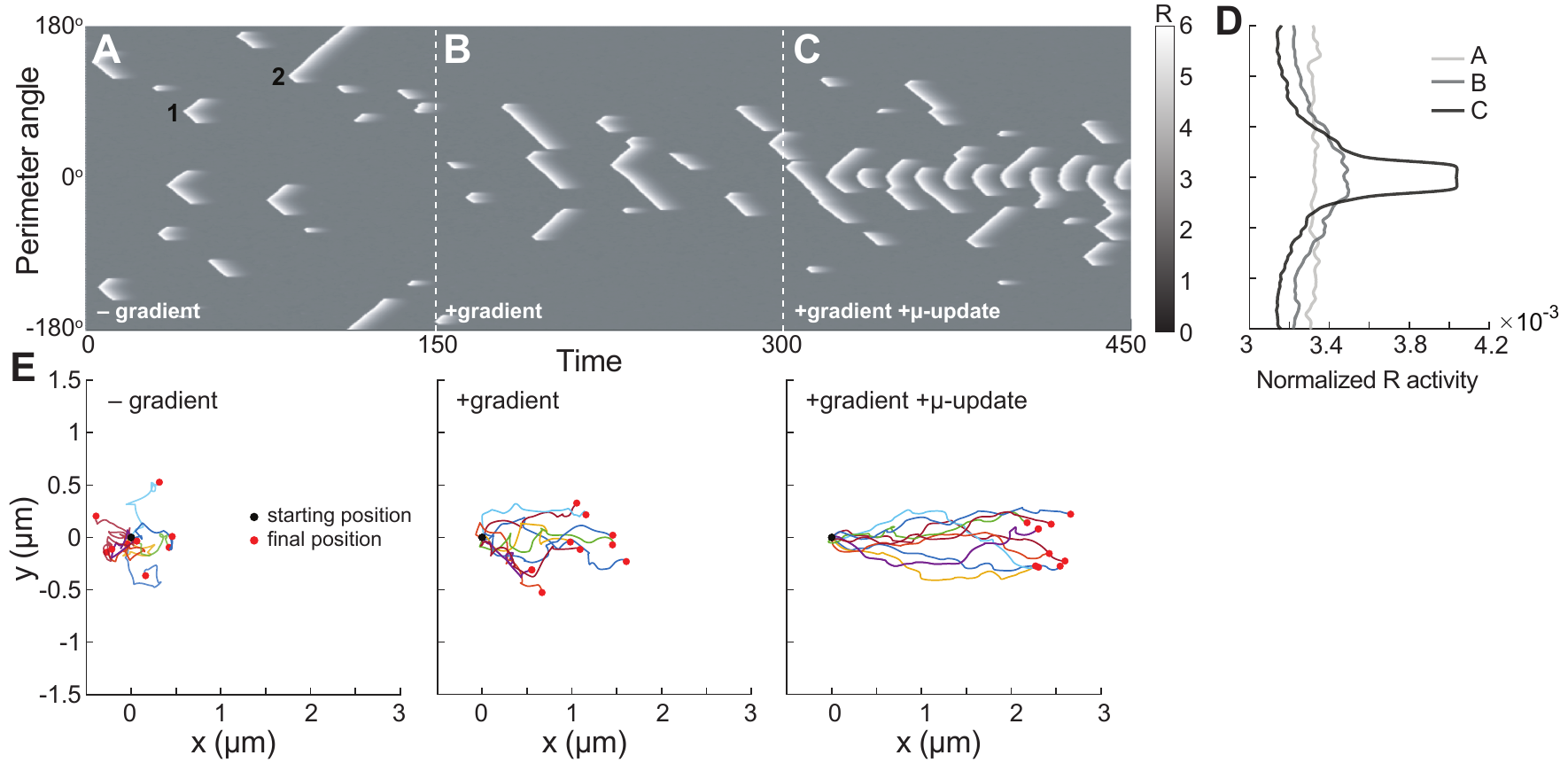}
\caption{\label{fig7}  %
Effect of $\mu$-update on directed migration. %
(A--C) Kymograph of $R$ showing the random firings (A,\,$t =0$--$150$), directed firings in presence of a gradient stimulus (B,\,$t =150$--$300$) and the effect of $\mu$-update law (C,\,$t =300$--$450$).  %
Random firings result in both symmetric (1) and asymmetric waves (2). %
The directional bias was introduced at $t= 150$ using \eref{eq:LEGI} with $\mu_1 = 0.32$. %
The $\mu$-update law was  spatially confined to the front of the cell by the Gaussian function $\exp(-\tfrac{1}{2}(\theta/\sigma)^2)$ with $\sigma = 0.2 (L/2\pi)^2$, $L = 30$. The diffusion coefficients chosen are: $D_F=0.25, D_R=0.15$. The rest of the parameters are as in \fref{fig5}. % 
(D) Total membrane activity of $R$ across the cell perimeter over a fixed time interval marking the three transitions for panels A (light gray), B (gray) and C(black), respectively.  The curves are normalized by the net activity in the region. %
(E) Trajectories of 10 cells undergoing random motility (left), directed migration (middle) and directed migration with $\mu$-adaptation for a period of 150 time units. The cells started at origin and the final positions are denoted with red circles. The viscoelastic model parameters are as in~\cite{Biswas:2022aa}.}
\end{figure}
%-------------------------------------------------------------------------------------------------

We simulated this system assuming no spatial heterogeneity and plotted the activity of $R$ as a function of time and space in \fref{fig7}A.  The system is initially at its stable equilibrium but is subject to noise.  

In this \emph{kymograph}, waves of high activity appear as v-shaped structures (wave marked 1, \fref{fig7}A).  This form appears because, following the noise-induced triggering of activity, symmetric waves move in opposite directions of the trigger point.  These waves have a natural life-time that is determined by the rise and diffusion of the inhibitory component $R$~\cite{Bhattacharya:2020aa}, in which case the two arms of the wave stop almost simultaneously. Alternatively,  one branch of the wave can cease to propagate because of stochastic effects \textcolor{black}{(wave marked 2, \fref{fig7}A)}.  Note that because of the absence of a spatial heterogeneity, the probability of triggering a wave is uniform over the perimeter.  

To determine our base-line chemotactic performance, we introduced a spatial gradient that changes the parameter $\mu$ along  the perimeter as:
\begin{align}
	\label{eq:LEGI}
	\mu=\mu_{i}-\mu_{1}\cos(\theta)
\end{align}
This is meant to recreate the effect of the signaling system that senses chemoattractant gradient and relays this to the excitable network.  \emph{Dictyostelium} cells employ an incoherent feedforward adaptation mechanism for altering the threshold, combining a fast excitation processes and slower inhibition~\cite{Parent:1999aa, Levchenko:2002aa, Takeda:2012aa, Tang:2014aa}.  This acts as a preprocessor that filters out the mean level of chemoattractant, allowing the cell to respond solely to the external spatial gradient.  Moreover, the fact that the excitation and inhibition processes represent local and global, respectively, receptor occupancy, means that the threshold is lowered at the front and raised at the rear of the cell, enabling efficient chemotaxis~\cite{Parent:1999aa, Levchenko:2002aa, Tang:2014aa}.   In our simulations, the effect of this local excitation, global inhibition scheme (LEGI) is obtained through the second term in \eref{eq:LEGI} where parameter $\mu_1$ controls its strength.  Simulation of the system incorporating \eref{eq:LEGI} is shown in  \fref{fig7}B starting at $t=150$.  We see that the waves of activity are confined to the region around zero degrees, representing the location of the gradient.   

We next sought to test the usefulness of the adaptation law \eref{eq:EN_adapt}.  However, for this to have an effect on chemotaxis, this adaptation must be done in a spatially-dependent manner. 
 Specifically, we activated  it in a narrow region of space:
 \begin{equation}
\label{eq:EN_adapt_2D}
	\dot{\mu}_i = e^{-\theta^2/2\sigma^2}\varepsilon_{\mu_i}\bigg(\dfrac{1}{1+\mathcal{C}^\star}-k\mu_{i}\bigg) 
\end{equation}
Thus, where the chemoattractant is highest, the value of $\mu$ adapts towards the bifurcation point; away from the chemoattracant gradient, $\mu_i$ remains unaltered. \fref{fig7}C shows the effect of this perturbation starting at $t=300$.  We see that the waves of activity are still confined to the region around zero degrees, representing the location of the gradient. More importantly, the number of waves increases greatly indicative of higher activity towards the gradient (\fref{fig7}D). 

To determine the effect of these firings on cell chemotaxis, we followed an approach that simulates the movement of cells
 using a  center-of-mass approximation~\cite{Biswas:2022aa}.  Using the spatially-depended level of activity, shown in \fref{fig7}A--C, we generated a series of force vectors normal to the cell surface.  The vector sum of all these vectors was used to obtain a  net protrusive force. After scaling this force so that it is in the range of experimentally observed protrusive pressures (0.5--5\,nN/$\mu\text{m}^2$), we use it to push a viscoelastic model of \emph{Dictyostelium} mechanics \cite{Yang:2008aa}.  In this model, the net stress in the $x$-direction: ${\sigma}_x$ (the direction of the gradient) alters the center-of-mass position ($\mathrm{CM}_x$) through the following dynamics:
\[
	\ddot{x}+({k_c}/{\gamma_c})\dot{x}
	=(1/\gamma_c+1/\gamma_a)\dot{{\sigma}}_x+(k_c/\tau_c^2){\sigma}_x,
\]
with a similar equation for the displacement in the $y$-direction ($\mathrm{CM}_y$). %

\textcolor{black}{
To illustrate the movement generated through this approximation, we first plotted the trajectories of 10 cells migrating randomly in the absence of any directional bias (\fref{fig7}E, left panel) and 
compared it to the simulated trajectories of equal number of cells in the presence of a gradient stimulus without  (\fref{fig7}E, middle panel) and with the adaptation law implemented (\fref{fig7}E, right panel).
 As expected, in the absence of any gradient, the center of mass exhibits a random walk.  When a gradient stimulus was applied,  the cells showed directed migration towards the right (source of chemoattractant).  Once the adaptation law was turned on, the trajectories were strongly directed to the right and in the same time duration, cells traveled almost twice the distance on average when compared to the former. 
}

%-------------------------------------------------------------------------------------------------
\section{Discussion}\label{sec6}
%-------------------------------------------------------------------------------------------------
Efficient chemotaxis requires that the signaling system be highly sensitive to small spatial heterogeneities in the concentration of the guiding chemoattractant.  The threshold of an excitable system provides an ideal mechanism for providing this sensitivity. It allows filtering of small stochastic fluctuations thus preventing cells from ``following the noise." On the other hand, it can greatly amplify persistent small differences in chemoattractant gradients.  A crucial requirement of such a highly nonlinear amplifier is that the set point be place near the edge of the threshold, but not at a point where the cell becomes oscillatory, as this actually hinders chemotaxis~\cite{Miao:2017aa}.   Even worse, moving the operating point beyond the second bifurcation point leads to cells that are permanently active~\cite{Edwards:2018aa}. These pancake-looking cells are so stretched and thin that they eventually die through 
fragmentation. 

 While the corresponding local excitation, global inhibition (LEGI) can work efficiently to guide cell movement, it still requires that the level of the LEGI preprocessor be tuned to the threshold that follows; without this matching, the gradient sensing mechanism is not robust~\cite{Levchenko:2002aa}. 
Here we have shown that an adaptation law, as suggested by Moreau et al.~\cite{Moreau:2003aa,Moreau:2003ab} would allow this matching of the steady-state response of the system with the threshold of the excitable system.  

It is worth asking whether this adaptation law would be implementable.  We highlight three hurdles. The first is that we require a reasonably good estimate of the function $f(x,y)$, or its equivalent in a more comprehensive model of the signaling network.  As shown in \fref{fig1}C, the wrong choice can move the system far from the equilibrium.  However, as suggested by how well the approximating ellipses work (e.g. \fref{fig4} and \fref{fig5}) it is unlikely that the desired trajectory needs to be specified with great accuracy.  This is particularly true when  one considers that that adaptation law that we are considering affects the mean level of the threshold. When combined with a mechanism for responding to the external gradient that lowers the threshold at the front and raises it at the rear, the adaptation law would still likely increase the chemotactic efficiency. The robustness of this scheme is an area for future consideration.

Second, the scheme relies on the possibility of implementing a complicated formula such as \eref{eq:EN_adapt} using biochemical components.  It should be pointed out that the actual form of $\mathcal{C}^\star$ follows the general form of many types of enzyme inhibitors.  Moreover, there have been reports in the synthetic biology community about the means of using chemical reaction networks to compute either arbitrary polynomials~\cite{Salehi:2017aa} or even logarithms~\cite{Chou:2017uy}. Thus, the adaptation laws that we propose are not beyond the realm of possibility.

Third, we note that in the particular scheme that we propose to improve chemotactic efficiency, we seek to use the adaptation law to adjust the operating point close to the bifurcation point only at the front of the cell. During chemotaxis, spatial self-organization of  different biochemical species is observed where some proteins and phospholipids  localize to the front of the cell and others go to the back~\cite{Devreotes:2017aa}. This suggest several potential ways to implement the $\mu$-adaptation in a spatially selective manner. We discussed one method in Section~\ref{sec:chemotax}, where it was controlled by the gradient stimulus which could be also contributed by any ``front" species. As an alternative, the update law could be controlled by ``back" entities where higher concentration of back molecules represses both the reactions in~\eref{eq:FRdiff}. Even further with the back molecules we can implement a more complex adaptation law where different spatial sections of the cells get adapted to different $\mu$ levels.

Finally, we note that an alternative means of improving efficient chemotaxis is by altering the noise properties of the system. The firings of an excitable system occur as noise causes the system to cross the threshold periodically. This depends on the size of the threshold, but also on the variance of the noise. Through a technique known as \emph{absolute concentration robustness} (ACR) \cite{Anderson:2014aa}, the stochastic fluctuations of a biochemical network can be reduced. If this is done in a spatially-dependent manner so that the rear of the cell has smaller fluctuations and hence fewer firings, chemotactic efficiency is also improved~\cite{Biswas:2022aa}. Interestingly, we found that the ACR motif that would accomplish this is similar to some of the cell's signaling network, suggesting the possibility that the cell is already using some form of ACR.  Another aspect of future research would be to ask whether cells already employ an adaptation scheme similar to that proposed here.

%-------------------------------------------------------------------------------------------------
\section*{Acknowledgments}
%-------------------------------------------------------------------------------------------------
We thank members of the Iglesias lab for useful conversations, particularly Sayak Bhattacharya.
PAI also wishes to thank Eduardo Sontag for many years of interesting and fruitful discussions. It is an honor to consider him a colleague.
%-------------------------------------------------------------------------------------------------
\section*{Declarations}
%-------------------------------------------------------------------------------------------------
The authors declare that there are no competing interests.

\bibliographystyle{plos2015}
\bibliography{paibio}% common bib file

\begin{thebibliography}{10}

\bibitem{Guram:2019aa}
Guram K, Kim SS, Wu V, Sanders PD, Patel S, Schoenberger SP, et~al.
\newblock A Threshold Model for T-Cell Activation in the Era of Checkpoint
  Blockade Immunotherapy.
\newblock Front Immunol. 2019;10:491.

\bibitem{Bene:2020aa}
Bene L, Bagd{\'a}ny M, Damjanovich L.
\newblock Adaptive threshold-stochastic resonance (AT-SR) in MHC clusters on
  the cell surface.
\newblock Immunol Lett. 2020 01;217:65--71.

\bibitem{Narni-Mancinelli:2013aa}
Narni-Mancinelli E, Ugolini S, Vivier E.
\newblock Tuning the threshold of natural killer cell responses.
\newblock Curr Opin Immunol. 2013 Feb;25(1):53--8.

\bibitem{Hudspeth:2014aa}
Hudspeth AJ.
\newblock Integrating the active process of hair cells with cochlear function.
\newblock Nat Rev Neurosci. 2014 Sep;15(9):600--14.

\bibitem{Choe:1998un}
Choe Y, Magnasco MO, Hudspeth AJ.
\newblock A model for amplification of hair-bundle motion by cyclical binding
  of Ca2+ to mechanoelectrical-transduction channels.
\newblock Proc Natl Acad Sci U S A. 1998 Dec;95(26):15321--6.

\bibitem{Camalet:2000uy}
Camalet S, Duke T, J{\"u}licher F, Prost J.
\newblock Auditory sensitivity provided by self-tuned critical oscillations of
  hair cells.
\newblock Proc Natl Acad Sci U S A. 2000 Mar;97(7):3183--8.

\bibitem{Eguiluz:2000ui}
Egu{\'\i}luz VM, Ospeck M, Choe Y, Hudspeth AJ, Magnasco MO.
\newblock Essential nonlinearities in hearing.
\newblock Phys Rev Lett. 2000 May;84(22):5232--5.

\bibitem{Jackson:2019tj}
Jackson Z, Wiesenfeld K.
\newblock Dynamics of tinnitus and coordinated reset therapy.
\newblock Phys Rev E. 2019 May;99(5-1):052403.

\bibitem{Parent:1999aa}
Parent CA, Devreotes PN.
\newblock A cell's sense of direction.
\newblock Science. 1999 Apr;284(5415):765--70.

\bibitem{Haastert:2007aa}
van Haastert PJM, Postma M.
\newblock Biased random walk by stochastic fluctuations of
  chemoattractant-receptor interactions at the lower limit of detection.
\newblock Biophys J. 2007 Sep;93(5):1787--96.

\bibitem{Vicker:2002ab}
Vicker MG.
\newblock Eukaryotic cell locomotion depends on the propagation of
  self-organized reaction-diffusion waves and oscillations of actin filament
  assembly.
\newblock Exp Cell Res. 2002 Apr;275(1):54--66.

\bibitem{Xiong:2010aa}
Xiong Y, Huang CH, Iglesias PA, Devreotes PN.
\newblock Cells navigate with a local-excitation, global-inhibition-biased
  excitable network.
\newblock Proc Natl Acad Sci U S A. 2010 Oct;107(40):17079--17086.

\bibitem{Hodgkin:1948aa}
Hodgkin AL.
\newblock The local electric changes associated with repetitive action in a
  non-medullated axon.
\newblock J Physiol. 1948 Mar;107(2):165--81.

\bibitem{Bhattacharya:2018aa}
Bhattacharya S, Iglesias PA.
\newblock The threshold of an excitable system serves as a control mechanism
  for noise filtering during chemotaxis.
\newblock PLoS One. 2018;17(3):e0201283.

\bibitem{Miao:2017aa}
Miao Y, Bhattacharya S, Edwards M, Cai H, Inoue T, Iglesias PA, et~al.
\newblock Altering the threshold of an excitable signal transduction network
  changes cell migratory modes.
\newblock Nat Cell Biol. 2017;19(4):329--340.

\bibitem{Zhan:2020aa}
Zhan H, Bhattacharya S, Cai H, Iglesias PA, Huang CH, Devreotes PN.
\newblock An excitable {Ras/PI3K/ERK} signaling network controls migration and
  oncogenic transformation in epithelial cells.
\newblock Dev Cell. 2020 09;54(5):608--623.e5.

\bibitem{Westendorf:2013aa}
Westendorf C, Negrete J Jr, Bae AJ, Sandmann R, Bodenschatz E, Beta C.
\newblock Actin cytoskeleton of chemotactic amoebae operates close to the onset
  of oscillations.
\newblock Proc Natl Acad Sci U S A. 2013 Mar;110(10):3853--3858.

\bibitem{Huang:2013aa}
Huang CH, Tang M, Shi C, Iglesias PA, Devreotes PN.
\newblock An excitable signal integrator couples to an idling cytoskeletal
  oscillator to drive cell migration.
\newblock Nat Cell Biol. 2013 Nov;15(11):1307--1316.

\bibitem{Miao:2019aa}
Miao Y, Bhattacharya S, Banerjee T, Abubaker-Sharif B, Long Y, Inoue T, et~al.
\newblock Wave patterns organize cellular protrusions and control cortical
  dynamics.
\newblock Mol Syst Biol. 2019;15(3):e8585.

\bibitem{Moreau:2003aa}
Moreau L, Sontag E.
\newblock Balancing at the border of instability.
\newblock Phys Rev E Stat Nonlin Soft Matter Phys. 2003 Aug;68(2 Pt 1):020901.

\bibitem{Moreau:2003ab}
Moreau L, Sontag E, Arcak M.
\newblock Feedback tuning of bifurcations.
\newblock Syst \& Cont Lett. 2003;50:229--239.

\bibitem{FitzHugh:1961aa}
Fitz{H}ugh R.
\newblock Impulses and physiological states in theoretical models of nerve
  membrane.
\newblock Biophys J. 1961 Jul;1(6):445--466.

\bibitem{Gill:1978aa}
Gill PE, Murray W.
\newblock Algorithms for the solution of the nonlinear least-squares problem.
\newblock SIAM J Numer Anal. 1978;15(5):977--992.

\bibitem{Shi:2013aa}
Shi C, Huang CH, Devreotes PN, Iglesias PA.
\newblock Interaction of motility, directional sensing, and polarity modules
  recreates the behaviors of chemotaxing cells.
\newblock PLoS Comput Biol. 2013;9(7):e1003122.

\bibitem{Biswas:2021ab}
Biswas D, Devreotes PN, Iglesias PA.
\newblock Three-dimensional stochastic simulation of chemoattractant-mediated
  excitability in cells.
\newblock PLoS Comput Biol. 2021 07;17(7):e1008803.

\bibitem{Lindner:2004aa}
Lindner B, Garcia-Ojalvo J, Neiman A, Schimansky-Geier L.
\newblock Effects of noise in excitable systems.
\newblock Phys Rep-Rev Sec Phys Lett. 2004 March;392(6):321--424.

\bibitem{Biswas:2022aa}
Biswas D, Bhattacharya S, Iglesias PA.
\newblock Enhanced chemotaxis through spatially regulated absolute
  concentration robustness.
\newblock Int J Robust Nonlin. 2022;p. 1--22.

\bibitem{Bhattacharya:2020aa}
Bhattacharya S, Banerjee T, Miao Y, Zhan H, Devreotes PN, Iglesias PA.
\newblock Traveling and standing waves mediate pattern formation in cellular
  protrusions.
\newblock Science Advances. 2020;6(32).

\bibitem{Levchenko:2002aa}
Levchenko A, Iglesias PA.
\newblock Models of eukaryotic gradient sensing: {A}pplication to chemotaxis of
  amoebae and neutrophils.
\newblock Biophys J. 2002 Jan;82(1 Pt 1):50--63.

\bibitem{Takeda:2012aa}
Takeda K, Shao D, Adler M, Charest PG, Loomis WF, Levine H, et~al.
\newblock Incoherent feedforward control governs adaptation of activated {Ras}
  in a eukaryotic chemotaxis pathway.
\newblock Sci Signal. 2012 Jan;5(205):ra2.

\bibitem{Tang:2014aa}
Tang M, Wang M, Shi C, Iglesias PA, Devreotes PN, Huang CH.
\newblock Evolutionarily conserved coupling of adaptive and excitable networks
  mediates eukaryotic chemotaxis.
\newblock Nat Commun. 2014;5:5175.

\bibitem{Yang:2008aa}
Yang L, Effler JC, Kutscher BL, Sullivan SE, Robinson DN, Iglesias PA.
\newblock Modeling cellular deformations using the level set formalism.
\newblock BMC Syst Biol. 2008;2:68.

\bibitem{Edwards:2018aa}
Edwards M, Cai H, Abubaker-Sharif B, Long Y, Lampert TJ, Devreotes PN.
\newblock Insight from the maximal activation of the signal transduction
  excitable network in {D}ictyostelium discoideum.
\newblock Proc Natl Acad Sci U S A. 2018;115(16):E3722--E3730.

\bibitem{Salehi:2017aa}
Salehi SA, Parhi KK, Riedel MD.
\newblock Chemical Reaction Networks for Computing Polynomials.
\newblock ACS Synth Biol. 2017 01;6(1):76--83.

\bibitem{Chou:2017uy}
Chou CT.
\newblock Chemical reaction networks for computing logarithm.
\newblock Synth Biol (Oxf). 2017 Jan;2(1):ysx002.

\bibitem{Devreotes:2017aa}
Devreotes PN, Bhattacharya S, Edwards M, Iglesias PA, Lampert T, Miao Y.
\newblock Excitable Signal Transduction Networks in Directed Cell Migration.
\newblock Annu Rev Cell Dev Biol. 2017;33:103--125.

\bibitem{Anderson:2014aa}
Anderson DF, Enciso GA, Johnston MD.
\newblock Stochastic analysis of biochemical reaction networks with absolute
  concentration robustness.
\newblock J R Soc Interface. 2014 Apr;11(93):20130943.

\end{thebibliography}


\end{document}